\documentclass[%
 reprint,
 superscriptaddress,
 showpacs,preprintnumbers,
 amsmath,amssymb,
 aps,
pra,
 longbibliography,
lengthcheck,%
]{revtex4-1}

\usepackage{graphicx}
\usepackage{dcolumn}
\usepackage{bm}
\usepackage{color}
\usepackage{ulem}
\usepackage{hyperref}
\usepackage{braket}

\begin{document}

\preprint{APS/123-QED}

\title{
Anomalous Hall effect by chiral spin textures in two-dimensional Luttinger model
}

\author{Ryunosuke Terasawa}
\affiliation{
Department of Physics, Tokyo Institute of Technology, Meguro, Tokyo, 152-8551, Japan
}

\author{Hiroaki Ishizuka}
\affiliation{
Department of Physics, Tokyo Institute of Technology, Meguro, Tokyo, 152-8551, Japan
}

\date{\today}

\begin{abstract}
Long-range magnetic textures, such as magnetic skyrmion, give rise to rich transport properties in magnetic metals, such as the anomalous Hall effect related to spin chirality, a.k.a. topological Hall effect.
In addition to the topological Hall effect, recent studies on {\it non-centrosymmetric} magnets find that the spin-orbit interaction of itinerant electrons gives rise to novel contributions related to spin chirality, i.e., the chiral Hall effect.
In this work, we discuss that the spin-orbit interaction has a distinct, yet significant, effect on the anomalous Hall effect related to spin chirality in {\it centrosymmetric} magnets.
Using a scattering theory method, we find that the anomalous Hall effect related to scalar spin chirality in a two-dimensional Luttinger model is suppressed by more than one order of magnitude compared to the quadratic dispersion, and the contributions similar to the chiral Hall effect in Rashba model vanishes.
At the same time, a novel term that gives different Hall conductivity for the Bloch and Neel skyrmions occurs, thereby enabling the detection of the skyrmion helicity. 
The striking differences demonstrate the rich effect of crystal symmetry on the chirality-related anomalous Hall effect in materials with strong spin-orbit interaction.
\end{abstract}

\pacs{
}

\maketitle


\section{Introduction}
Noncollinear magnetic textures give rise to rich transport phenomena, such as anomalous~\cite{Ye1999,Ohgushi2000,Taguchi2001,Tatara2002,Nagaosa2010,Liu2018,Wang2018} and spin~\cite{Ishizuka2013,Ishizuka2021} Hall effects, and electrical magnetochiral effect~\cite{Yokouchi2017,Aoki2019,Ishizuka2020}.
Theoretically, these phenomena are often related to spin chirality: the anomalous Hall effect (AHE) is related to scalar spin chirality of three spins $\bm{S}_i\cdot\bm{S}_j\times\bm{S}_k$~\cite{Ye1999,Tatara2002,Ishizuka2018a} whereas the spin Hall effect~\cite{Ishizuka2021} and electrical magnetochiral effects~\cite{Ishizuka2020} are related to the vector spin chirality of two spins $\bm{S}_i\times\bm{S}_j$.
These phenomena play an important role in the transport properties of ferromagnetic materials with long-range magnetic structures.
The anomalous Hall effect related to scalar spin chirality is intensively studied in materials hosting magnetic skyrmions, such as B20 compounds~\cite{Neubauer2009,Kanazawa2011,Yokouchi2014}, pyrochlore magnets~\cite{Taguchi2001,Machida2007}, and Mn$_3$Sn~\cite{Nakatsuji2015}.
On the other hand, the electrical magnetochiral effect is reported in magnets with helical magnetic order~\cite{Yokouchi2017,Aoki2019}.
While most of the early works ignored the effect of spin-orbit interaction (SOI) of electronic bands, recent studies pointed out that the SOI gives rise to non-trivial contribution to the anomalous Hall effect by skyrmions~\cite{Lux2020,Yamaguchi2021,Mochida2022}, such as chiral Hall effect (CHE) and monopole contribution.
These studies revealed nontrivial effects of spin-orbit interaction in on non-centrosymmetric two-band models, such as Rashba and Dresselhaus models.

\begin{figure}[tbp]
  \includegraphics[width=\linewidth]{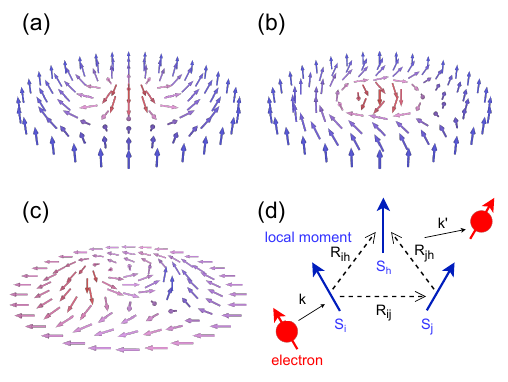}
  \caption{Schematics of (a) Neel- and (b) Bloch-type magnetic skyrmion and (c) Neel-type bimeron studied in this work. (d) A schematic of electron scattering by multiple local moments.
  }\label{fig:model}
\end{figure}

AHE by magnetic textures also occurs in centrosymmetric materials as skyrmion and magnetic helix also appear in centrosymmetric materials, such as in frustrated ~\cite{Okubo2012,Leonov2015} and itinerant magnets~\cite{Ozawa2017}.
These mechanisms not only broaden the list of candidate materials but are also favorable for realizing a large chirality-related AHE.
In the case of frustrated magnets, the size of the skyrmion stabilized by these mechanisms is governed by the ratio of competing exchange interactions, typically the nearest and further neighbor interactions, and by the Fermi wavenumber in the case of itinerant magnets.
In both cases, the typical size of a skyrmion is smaller than those stabilized by the Dzyaloshinskii-Moriya interaction in non-centrosymmetric materials, in which case the skyrmion radius $\lambda$ is related to the ratio of ferromagnetic exchange and Dzyaloshinskii-Moriya interaction, $\lambda=10-100$ nm~\cite{Tokura2013}. 
As the magnitude of AHE by skyrmions is proportional to the density of the skyrmion~\cite{Tokura2013}, a smaller skyrmion radius is favorable for realizing a larger AHE.
In fact, some centrosymmetric skyrmion materials show a very large topological Hall effect~\cite{Kurumaji2019}.
As with non-centrosymmetric materials, SOI may affect the chirality-related AHE in centrosymmetric materials.
However, while several works studies the effects of spin-orbit interaction on the AHE in coplanar magnets~\cite{Chen2014,Zhang2020}, the role of inversion symmetry remains unclear.

In this work, we theoretically study the AHE by magnetic textures in a centrosymmetric magnet with spin-orbit interaction.
Considering the two-dimensional Luttinger model~\cite{Luttinger1956} as the effective Hamiltonian for itinerant electrons, we study the AHE induced by long-range magnetic textures using a scattering theory method.
We show that the anomalous Hall effect related to scalar spin chirality is suppressed by more than one order of magnitude compared to the simple quadratic Hamiltonian without SOI, and the novel contribution discovered in the Rashba model vanishes.
On the other hand, a different contribution related to vector spin chirality appears that gives rise to the difference in the Hall conductivity in Bloch and Neel skyrmion crystals.
Thereby enabling electrical distinction between the two types of skyrmions [Fig.~\ref{fig:helicitydep}].
The effect of SOI on AHE by bimerons is also discussed.

\section{Model and Method}

\subsection{Luttinger Kondo-lattice model}
\begin{figure}
    \centering
    \includegraphics[width=\linewidth]{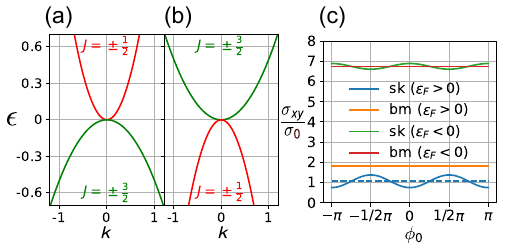}
    \caption{The band structure of the Luttinger model with $\alpha=0.5$, (a) $m=1/2$ and (b) $m=-1/2$ . (c) Helicity dependence of chirality-related anomalous Hall conductivity in skyrmion and bimeron crystals with $M_z=\alpha=0, S=n_{sk}=q=1$. In the figure, sk and bm denote skyrmion and bimeron, respectively. }
    \label{fig:helicitydep}
\end{figure}

As an example of the centrosymmetric semiconductor with SOI, we consider a two-dimensional variant of the Luttinger model coupled to classical spins.
The Hamiltonian reads 
\begin{align}
  H=H_0+H_K,
\end{align}
where, 
\begin{align}\label{Luttinger}
  &H_0=\nonumber\\
  &\sum_{\bm{k},\mu,\nu}c_{\mu}(\bm{k})^\dagger\frac{1}{2m}\left[\left(\frac{5}{4}+\alpha\right)k^2-(k_x J_x+k_y J_y)^2\right]_{\mu\nu}c_{\nu}(\bm{k})
\end{align}
is the effective Hamiltonian for itinerant electrons described by the Luttinger Hamiltonian~\cite{Luttinger1956} and
\begin{align}
  H_K=J_K\sum_ic_{\mu}^\dagger(\bm{R}_i)(\bm{S}_{i}\cdot\bm{J})_{\mu\nu}c_{\nu}(\bm{R}_i)
\end{align}
is the Kondo coupling between the classical localized spins and itinerant electrons.
Here, $c_\mu(\bm k)$ [$c_\mu^\dagger(\bm k)$] is the annihilation [creation] operator of an electron with momentum $\bm{k}=(k_x,k_y)$ and spin $\mu$, $k=|\bm{k}|$, $\bm{J}=(J_x,J_y,J_z)$  is the vector of $J=3/2$ spin operators $J_a$ ($a=x,y,z$),
$\bm{S}_{i}=(S_i^x,S_i^y,S_i^z)$ is the $i$th localized moment at position $\bm{R}_i=(R_i^x,R_i^y)$, and $J_K$ is the strength of exchange coupling between the itinerant electrons and the localized moment.
The eigenstates of $H_0$ consist of two doubly-degenerate bands [Fig.~\ref{fig:helicitydep}(a)].
When $|\alpha|<1$, one of the two doubly-degenerate bands is hole-like, and the other becomes electron-like;
the electron and hole bands touch at $\bm{k}=0$, forming a zero-gap semiconductor state.
Such a state is realized in $\alpha$-Sn~\cite{Zhang2018}, and in pyrochlore iridates~\cite{Moon2013,Rhim2015}.
On the other hand, when $|\alpha|>1$, both bands are either electron-like ($m\alpha>0$) or hole-like ($m\alpha<0$).
In this work, we focus on the $|\alpha|<1$ case, i.e., the case in which one of the two doubly-degenerate bands is electron-like, and the other is hole-like [Figs.~\ref{fig:helicitydep}(a) and ~\ref{fig:helicitydep}(b)].

\subsection{Anomalous Hall effect by skew scattering}
To study the AHE arising from coupling to magnetic textures, we compute the anomalous Hall conductivity $\sigma_{xy}$ focusing on the skew scattering contribution~\cite{Smit1955}.
In the Boltzmann theory, the skew scattering is described by the asymmetry of the scattering rate.
The scattering rate $W_{\bm{k}\mu\rightarrow \bm{k}'\nu}$ is the rate of electron in $\Ket{\bm k\mu}$ state being scattered to the state $\Ket{\bm k'\nu}$.
The skew scattering AHE is related to the difference of $W_{\bm{k}\mu\rightarrow \bm{k}'\nu}$ and its inverse process $W_{\bm{k}'\nu\rightarrow \bm{k}\mu}$. 
To study the asymmetry in the scattering rate, we first define the symmetric $w_{\bm{k}\mu\rightarrow \bm{k}'\nu}^+$ and antisymmetric $w_{\bm{k}\mu\rightarrow \bm{k}'\nu}^-$ terms of the scattering rate by 
\begin{align}
  w_{\bm{k}\mu\rightarrow \bm{k}'\nu}^{\pm}=\frac{1}{2}(W_{\bm{k}\mu\rightarrow \bm{k}'\nu}\pm W_{\bm{k}'\nu\rightarrow \bm{k}\mu}).
\end{align}
In this work, we calculate $w_{\bm{k}\mu\rightarrow \bm{k}'\nu}^-$ by the magnetic scattering to study the AHE.

Within the second Born approximation, the antisymmetric term $w_{\bm{k}\mu\rightarrow \bm{k}'\nu}^-$ reads
\begin{align}\label{eq:skew}
      &w_{\bm{k}\mu\rightarrow \bm{k}'\nu}^- = 4\pi^2\sum_{\bm{p},\lambda} \Im\left[\Braket{\bm{k}'\nu|H_K|\bm{k}\mu}\Braket{\bm{k}\mu|H_K|\bm{p}\lambda}\right.\nonumber\\
  &\left.\times\Braket{\bm{p}\lambda|H_K|\bm{k}'\nu}\right] \delta (\epsilon_{\bm{k}\mu} - \epsilon_{\bm{k}'\nu} ) \delta(\epsilon_{\bm{k}\mu} - \epsilon_{\bm{p}\lambda} ).
\end{align}
Here, we considered $H_0$ as the unperturbed Hamiltonian and $H_K$ as the perturbation that causes electron scattering.
The Hall conductivity is calculated by combining Eq.~\eqref{eq:skew} with the semiclassical Boltzmann theory, as summarized in the supplemental information~\cite{Suppl}.

\section{Results}
\subsection{Hall conductivity for the electron-doped case}
We first consider electron-doped case $(\epsilon_F\geq0)$ assuming a positive electron mass $m>0$, where $\epsilon_F$ is the Fermi energy.
The Hall conductivity $\sigma_{xy}$ reads
\begin{align}
    &\sigma_{xy}=\nonumber\\
    &-\frac{\sigma_{0}}{2^{10}(1+\alpha)}\left[9f_1(\{\bm{S}_{h}\})+58f_2(\{\bm{S}_{h}\})+30f_3(\{\bm{S}_{h}\})\right].\label{eq:sxy-elec}
\end{align}
Here,
\begin{align*}
    f_1&(\{\bm{S}_{h}\})=\frac{1}{L^2}\sum_{h,i,j}S_h^z[8S_i^zS_j^z\bm{R}_{ih}\cdot\bm{R}_{jh}\nonumber\\
    &+(\bm{S}_{i}\cdot \bm{S}_{j})(17\bm{R}_{ih}\cdot\bm{R}_{jh}-5R_{ih}^2-5R_{jh}^2)],\\
  f_2&(\{\bm{S}_{h}\})=\frac1{L^2}\sum_{h,i,j}(\bm{S}_{h}\cdot\bm{S}_{i}\times \bm{S}_{j})(\bm{R}_{ih}\times \bm{R}_{jh}\cdot \hat{\bm{z}}),\\
    f_3&(\{\bm{S}_{h}\})=\nonumber\\
    &\frac1{L^2}\sum_{h,i,j}(\bm{S}_{h}\cdot\bm{R}_{ij})\left[\bm{S}_{i}\times \bm{S}_{j}\cdot(\bm{R}_{ih}+\bm{R}_{jh})\times\hat{\bm{z}}\right]\nonumber\\
    &+(\bm{S}_{h}\cdot\bm{R}_{ij}\times\hat{\bm{z}})\left[\bm{S}_{i}\times \bm{S}_{j}\cdot(\bm{R}_{ih}+ \bm{R}_{jh})\right],
\end{align*}
and $\sigma_0=\frac{\tau^2e^2|m|^3J_K^3\epsilon_F^2}{2\pi}$, where $\tau$ is the relaxation time for symmetric scattering and $e$ is electric charge.
Among three terms, $f_1(\{\bm{S}_{h}\})$ corresponds to a generalization of the skew scattering AHE by magnetic scattering~\cite{Kondo1962} and $f_2(\{\bm{S}_{h}\})$ is the skew scattering AHE related to the scalar spin chirality $\bm{S}_{h}\cdot\bm{S}_{i}\times \bm{S}_{j}$~\cite{Ishizuka2018a}.
The third term $f_3(\{\bm{S}_{h}\})$ is the AHE related to the vector spin chirality $\bm{S}_{i}\times \bm{S}_{j}$, which is the term describing the interplay of magnetic texture and SOI. 

To see how the SOI affects the AHE, we compare Eq.~\eqref{eq:sxy-elec} to the anomalous Hall conductivity of a model without SOI.
To this end, we consider a quadratic Hamiltonian
\begin{align}
    \tilde H_0=\sum_{\bm{k}}c_{\mu}(\bm{k})^\dagger\frac{(1+\alpha)k^2}{2m}c_{\mu}(\bm{k}).
\end{align}
The band structure of $\tilde H_0$ is exactly the same as the electron band of $H_0$.
Therefore, comparing the anomalous Hall conductivity helps us clarify the effect of SOI.
The scattering rate and Hall conductivity for $\tilde H_0$ reads
\begin{align}
  \tilde\sigma_{xy}=-\frac{\sigma_0}{1+\alpha}f_2(\{\bm S_h\}).
\end{align}
The result resembles the $f_2(\{\bm S_h\})$ term in $\sigma_{xy}$.
However, the magnitude in Eq.~\eqref{eq:sxy-elec} is reduced by $29/512\sim1/18$ compared to the case without SOI.
The result indicates that the AHE related to scalar spin chirality in the Luttinger model is suppressed by more than one order of magnitude compared to that without SOI.

\subsection{Continuum limit for the electron-doped case}
\begin{figure}[t]
    \centering
    \includegraphics{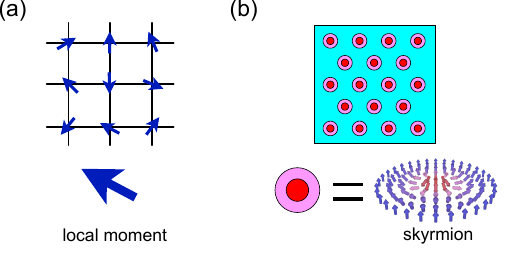}
    \caption{A schematic of (a) local spins on the square lattice and (b) an example of skyrmion crystal.}
    \label{fig:square}
\end{figure}

We next turn to a spin texture slowly varying in space, such as in ferromagnets with magnetic skyrmions.
To be concrete, we consider a square lattice magnet [Fig.~\ref{fig:square}(a)].
When the spins vary slowly in space, $\bm{S}_i$ near $\bm S_h$ can be approximated as $\bm{S}_{i}\sim\bm{S}_{h}+(\bm{R}_{ih}\cdot\nabla)\bm{S}_{h}+\frac{1}{2}(\bm{R}_{ih}\cdot\nabla)^2\bm{S}_{h}$.
To evaluate the Hall conductivity, we assume that the contribution from the multiple-spin scattering due to nearest-neighbor sites is dominant.
That is, we limit $i$ and $j$ to the nearest-neighbor sites of $h$.
By using the gradient expansion and replacing $\sum_h$ by $\int \frac{dxdy}{a^2}$, we obtain 
\begin{widetext}
\begin{align}
  &\sigma_{xy}^{(h)}=\frac{\sigma_0}{(1+\alpha)L^2}\int dxdy\;\frac{45}{32}S^zS^2-\frac{153}{256}a^2S^z(|\partial_x\bm{S}|^2+|\partial_y\bm{S}|^2)+\frac{45}{64}a^2S^z\bm{S}\cdot\Delta\bm{S}-\frac{9}{32}a^2S^z(|\partial_xS^z|^2+|\partial_yS^z|^2)\nonumber\\
  &-\frac{29}{64}\frac{\sigma_0 a^2}{(1+\alpha)L^2}\int dxdy\;\bm{S}\cdot\partial_x\bm{S}\times\partial_y\bm{S}+\frac{15}{128}\frac{\sigma_0 a^2}{(1+\alpha)L^2}\int dxdy\;(S^x[(\partial_x^2-\partial_y^2)\bm{S}\times \bm{S}]_y+S^y[(\partial_x^2-\partial_y^2)\bm{S}\times \bm{S}]_x).\label{eq:sxy-cont}
\end{align}
\end{widetext}

Compared to non-centrosymmetric models~\cite{Lux2020,Mochida2022}, no term with one spatial-derivative of spin exists in Eq.~\eqref{eq:sxy-cont} such as the contributions similar to CHE~\cite{Lux2020} and monopole contribution are absent~\cite{Mochida2022}.
The absence of linear-in-gradient terms is understandable from the symmetry of $H$.
Phenomenologically, Eq.~\eqref{eq:sxy-cont} indicates that the Hall current follows $J_y=\sigma' f(\{\bm S(\bm r)\})E_x$, where $f(\{\bm S(\bm r)\})$ is a functional of spins.
For concreteness, let us consider $f(\{\bm S(\bm r)\})=(S^x)^2\partial_x S^x$.
In this case, the inversion operation transforms $J_y=\sigma' f(\{\bm S(\bm r)\})E_x$ $\to$ $J_y=-\sigma' f(\{\bm S(\bm r)\})E_x$. Hence, $\sigma'=0$.
The same argument holds for arbitrary $f(\{\bm S(\bm r)\})$ with one spatial derivative.
Hence, the AHE related to terms with one spatial derivative, e.g., CHE and monopole contribution, is prohibited in a centrosymmetric system.
Instead, two-derivative terms, such as $S^x[(-\partial_x^2+\partial_y^2)\bm{S}\times \bm{S}]_y+S^y[(-\partial_x^2+\partial_y^2)\bm{S}\times \bm{S}]_x$ and $\bm{S}\cdot\partial_x\bm{S}\times\partial_y\bm{S}$, are allowed by symmetry.
Therefore, the absence of the CHE and monopole contribution reflects the inversion symmetry of the model.

\subsection{Magnetic skyrmion and bimeron}
To gain insight into how the novel term affects AHE in skyrmion materials, we apply Eq.~\eqref{eq:sxy-cont} to a skyrmion crystal state.
Here, we consider a crystal of a magnetic skyrmion whose spin configuration is
\begin{align}\label{eq:skyrmion}
  \bm{S}=S\left(\frac{2\lambda r\cos(q\phi-\phi_0)}{r^2+\lambda^2},\frac{2\lambda r\sin(q\phi-\phi_0)}{r^2+\lambda^2},\frac{r^2-\lambda^2}{r^2+\lambda^2}\right),
\end{align}
where $r$ is the distance from the skyrmion center, $\lambda$ is the skyrmion radius, $\phi$ is the azimuth angle ~\cite{Belavin1975}, $\phi_0$ is the helicity of the skyrmion, and $q=\pm1$ is the skyrmion charge~\cite{Tokura2013}.
When $\phi_0=0, \pi$, the skyrmion is Ne\'el type, and when $\phi_0=\pm\pi/2$, it is Bloch type [See Figs.~\ref{fig:model}(a) and \ref{fig:model}(b)].
A periodic alignment, or a crystal, of skyrmions [see Fig.~\ref{fig:square}(b)] is known to be stable in many magnetic materials~\cite{Muelbauer2009,Yu2010}, which has been intensively explored over the last couple of decades~\cite{Kézsmárki2015,Tokura2021,Saha2022}.

By using Eq.~\eqref{eq:sxy-cont}, the anomalous Hall conductivity of the skyrmion crystal reads
\begin{align}
  \sigma_{xy}&=\frac{\sigma_0 S^3}{32(1+\alpha)}\left[45M_z+n_{sk}[34q-10\cos(2\phi_0)]\right],
\end{align}
where 
$M_z$ is the out-of-plane magnetization per area and $n_{sk}$ is the skyrmion density.
The term proportional to $n_{sk}q$ contains the contribution from $S^z(|\partial_xS^z|^2+|\partial_yS^z|^2)$ and scalar spin chirality $\bm{S}\cdot\partial_x\bm{S}\times\partial_y\bm{S}$, whereas the third term proportional to $n_{sk}\cos(2\phi_0)$ is the novel contribution from SOI.
The $\cos(2\phi_0)$ dependence of the third term indicates that the Hall conductivity depends on the helicity; the term gives a negative contribution to AHE in Neel-type skyrmion and a positive contribution to Bloch type.
Hence, the $f_3$ term, which arises from the interplay of SOI and magnetic texture, enables the detection of the skyrmion helicity. 

As another example of nontrivial magnetic texture, we consider bimeron crystal whose spin orientation reads [Fig.~\ref{fig:model}(c)]
\begin{align}\label{eq:bimeron}
  \bm{S}=S\left(\frac{r^2-\lambda^2}{r^2+\lambda^2},\frac{2\lambda r\sin(q\phi-\phi_0)}{r^2+\lambda^2},-\frac{2\lambda r\cos(q\phi-\phi_0)}{r^2+\lambda^2}\right).
\end{align}
Intuitively, magnetic bimeron is a skyrmion whose spins are rotated by $\pi/2$ about $x$ or $y$ axis ~\cite{Nagase2021}.
The anomalous Hall conductivity for the bimeron crystal reads
\begin{align}
  \sigma_{xy}&=\frac{29}{16}\frac{\sigma_0}{1+\alpha}n_{sk}qS^3.
\end{align}
Unlike the magnetic skyrmion case, only $f_2(\{\bm S_h\})$ term contributes to the AHE in the bimeron crystal.


\subsection{Hole-doped case}
We next consider the hole-doped case ($\epsilon_F \leq0 $), assuming $m>0$. The Hall conductivity reads
\begin{align}
\sigma_{xy}=\frac{27\sigma_0}{2^{9}(1-\alpha)}\left[f'_1(\{\bm{S}_{h}\})-4f_2(\{\bm{S}_{h}\})+\frac14f_3(\{\bm{S}_{h}\})\right],\label{eq:sxy-hole}
\end{align}
where
\begin{align*}
    f'_1&(\{\bm{S}_{h}\})=\nonumber\\
    &\frac1{L^2}\sum_{h,i,j}S_h^z[(\bm{S}_{i}\cdot \bm{S}_{j})(R_{ih}^2+R_{jh}^2-6\bm{R}_{ih}\cdot\bm{R}_{jh})].
\end{align*}
In the continuum limit, the Hall conductivity reads
\begin{widetext}
\begin{align}
  &\sigma_{xy}=\frac{\sigma_0}{(1-\alpha)L^2}\int dxdy\;\frac{27}{16}S^zS^2-\frac{81}{64}a^2S^z(|\partial_x\bm{S}|^2+|\partial_y\bm{S}|^2)+\frac{27}{32}a^2S^z\bm{S}\cdot\Delta\bm{S}\nonumber\\
  &-\frac{27\sigma_0}{16(1-\alpha)L^2}\int dxdy\;\bm{S}\cdot\partial_x\bm{S}\times\partial_y\bm{S}-\frac{27\sigma_0}{512(1-\alpha)L^2}\int dxdy\;(S^x[(\partial_x^2-\partial_y^2)\bm{S}\times \bm{S}]_y+S^y[(\partial_x^2-\partial_y^2)\bm{S}\times \bm{S}]_x).\label{eq:sxy-hole-cont}
\end{align}
\end{widetext}
Compared to the electron-doped case, the $f_2$ term becomes relatively large compared to $f_3$, i.e., the topological Hall effect becomes larger compared to the effect of SOI in the hole-doped case.

By using this equation, the anomalous Hall conductivity of the skyrmion crystal reads 
\begin{align}
    \sigma_{xy}=\frac{27\sigma_0S^3}{32(1-\alpha)}\left[2M_z+n_{sk}\left(8q+\frac{\cos(2\phi_0)}{6}\right)\right],
\end{align}
and the anomalous Hall conductivity of the bimeron crystal reads
\begin{align}
    \sigma_{xy}=\frac{27}4\frac{\sigma_0}{1-\alpha}n_{sk}qS^3.
\end{align}
As expected from Eq.~\eqref{eq:sxy-hole}, the helicity dependence of $\sigma_{xy}$ for the skyrmion crystal becomes smaller compared to the electron-doped case and with the opposite sign, as shown in Fig.~\ref{fig:helicitydep}(c).

\subsection{$m<0$ case}
When $m<0$, the same calculation gives Eqs.~\eqref{eq:sxy-hole} and \eqref{eq:sxy-hole-cont} for $\varepsilon_F>0$ and Eqs.~\eqref{eq:sxy-elec} and \eqref{eq:sxy-cont} for $\varepsilon_F<0$, i.e., the result for electron- and hole-doped cases inverts [Fig.~\ref{fig:helicitydep}(a)].
This is because the orbitals for electron and hole bands invert by changing the sign of $m$.
Hence, the behavior of AHE in the $m<0$ case is qualitatively the same as what we discussed above.

Note that this behavior is distinct from the Rashba model~\cite{Lux2020,Mochida2022}, where changing the sign of spin-orbit interaction changes the sign of AHE.
For the case of Rashba model, inverting the sign of spin-orbit interaction only changes the spin texture in the momentum space; the electron dispersion remains exactly the same.
On the other hand, the four bands in the Luttinger model are usually the bands that split off from 6 or more degenerated bands in the absence of spin-orbit interaction~\cite{Luttinger1955a}.
In such a case, changing the sign of spin-orbit interaction inverts the orbitals, leading to a behavior more complicated than the Rashba model.
Hence, the AHE in Luttinger model shows distinct behavior compared to the Rashba model.

\section{Summary}
In this work, we studied the skew scattering and anomalous Hall effect in a two-dimensional Luttinger model coupled to localized moments by the Kondo coupling.
Using a scattering theory method, we derive a general formula for the anomalous Hall conductivity, which consists of three terms:
the AHE proportional to the magnetization, THE, and  vector spin chirality term.
The result shows that THE in the Luttinger model is suppressed by more than one order of magnitude compared to the quadratic band electron.
On the other hand, the vector spin chirality term is a contribution arising from the interplay of SOI and non-trivial magnetic texture.
This term, however, is a distinct term from those known in non-centrosymmetric models such as CHE related to vector spin chirality~\cite{Lux2020} and monopole contribution~\cite{Mochida2022}.
This is evident from the fact that only the second-order derivative of spins appears in the continuum limit, in contrast to the vector spin chirality $\bm S\times\nabla\bm S$ and magnetic monopole $\nabla\cdot\bm S$ terms in non-centrosymmetric magnets.
The novel term gives rise to the helicity dependence of anomalous Hall conductivity in the case of skyrmion crystal.
Hence, it enables the electrical distinction of Bloch and Neel skyrmions.
In the case of bimeron crystal, however, only THE term contributes to the AHE.
Hence, there is no helicity dependence. The results demonstrate the rich effect of SOI and crystal symmetry on the anomalous Hall effect by non-trivial spin texture.

\acknowledgements
We are grateful for the fruitful discussions with J. Fujii, N. Kanazawa, and J. Mochida.
This work was supported by JSPS KAKENHI (Grant Numbers JP19K14649 and JP23K03275).

\appendix
\section{Boltzmann theory}\label{sec:Boltzmann}

The calculation of Hall conductivity follows a paper by one of the authors~\cite{Ishizuka2018a}. In the Boltzmann theory, the electron distribution was evaluated by the Boltzmann equation. In the presence of the uniform static electric field in $x$ direction, the Boltzmann equation reads 
\begin{align}\label{eq:Boltzmann}
    &\frac{(1+\alpha)ek\cos\theta_kE_x}{m}f_0'(\epsilon_{\bm{k}})=\nonumber\\
    &\qquad\frac{g_{\bm{k}\mu}}{\tau}+\frac{L^2}{4\pi^2} \sum_\nu\int dk'd\theta_{k'}\; w_{\bm{k}\mu\rightarrow \bm{k}'\nu}^- g_{\bm{k}'\nu},
\end{align}
where $E_x$ is the external electric field, and $f_0(\epsilon_{\bm{k}})$ and $f_0'(\epsilon_{\bm{k}})$ are the Fermi-Dirac distribution function and its energy derivative, respectively. Here, we assumed that the electron occupation $f_{\bm{k}\mu}=f_0(\epsilon_{\bm{k}})+g_{\bm{k}\mu}$ is close to that of the Fermi-Dirac distribution and expanded to the equation up to leading order in $E_x$, assuming $g_{\bm{k}\mu}=\mathcal{O}(E_x)$. In addition, Eq.~\ref{eq:Boltzmann}, we used the relaxation time approximation for the symmetric part of the scattering rate $w_{\bm{k}\mu\rightarrow \bm{k}'\nu}^+$, i.e., $w_{\bm{k}\mu\rightarrow \bm{k}'\nu}^+$ is replaced by $-\frac{g_{\bm{k}\mu}}{\tau}$, where $\tau$ is the relaxation time. 

Here, we assume the form 
\begin{align}\label{eq:generic form}
    w_{\bm{k}\mu\rightarrow \bm{k}'\nu}^-=\sum_n c_n\sin n\phi,
\end{align}
where $\phi=\theta_{k'}-\theta_k$ is the angle between $\bm{k}$ and $\bm{k}'$. This is a generalization of the antisymmetric scattering term. Solving Eq.~\ref{eq:Boltzmann} up to leading order in $w_{\bm{k}\mu\rightarrow \bm{k}'\nu}^-$, $g_{\bm{k}\mu}$ reads
\begin{align}
    &g_{\bm{k}\mu}=\frac{(1+\alpha)\tau ek\cos\theta_kE_x}{m}f_0'(\epsilon_{\bm{k}})+\nonumber\\
    &\frac{(1+\alpha)\tau eL^2E_x}{4\pi^2m} \sum_{\nu,n} c_n \int k' f_0'(\epsilon_{\bm{k}'}) \;dk' \int \sin n\phi \cos\theta_{k'}\;d\theta_{k'}.
\end{align}
Since $\int \sin n\phi \cos\theta_{k'}\;d\theta_{k'}=-\frac{\sin \theta_k}{2} \delta_{n,1}$, only $c_1$ term in Eq.~\ref{eq:generic form} remains i.e., among the asymmetric scattering terms, only those proportional to $\sin\phi$ contribute to Hall conductivity.

\section{Anomalous Hall conductivity}

Using Eq.~(5) in the main text, the antisymmetric scattering term $w_{\bm{k}\mu\rightarrow \bm{k}'\nu}^-$ for hole-doped case reads, 
    \begin{align}
    w_{\bm{k}\mu\rightarrow \bm{k}'\nu}^-=\frac{27\pi |m| J_K^3 }{2^9(1-\alpha)L^4 }V_{\mu\nu}\delta\left(\epsilon_{\bm{k}\mu}-\epsilon_{\bm{k}'\nu}\right),
   \end{align}
where
\begin{align}
    \bm{V}=V_0\sigma_0+V_x\sigma_x+V_y\sigma_y+V_z\sigma_z,
\end{align}
and
\begin{widetext}
   \begin{align}
    V_0&=S_h^z(\bm{S}_{i}\cdot \bm{S}_{j})(\sin\phi+2\sin2\phi+\sin3\phi)\nonumber\\
    &+k^2S_h^z(\bm{S}_{i}\cdot \bm{S}_{j})\frac{1}{4}[R_{ij}^2(\sin\phi-4\sin2\phi-3\sin3\phi)+\bm{R}_{ih}\cdot\bm{R}_{jh}(-4\sin\phi+2\sin2\phi+4\sin3\phi+\sin4\phi)]\nonumber\\
    &+k^2(\bm{R}_{ih}\times \bm{R}_{jh}\cdot \hat{z})[\bm{S}_{h}\cdot\bm{S}_{i}\times \bm{S}_{j}(2\sin\phi+\sin2\phi)+S_h^z(\bm{S}_{i}\times \bm{S}_{j}\cdot \hat{z})\frac{1}{4}(-4\sin\phi+2\sin2\phi+4\sin3\phi+\sin4\phi)]\nonumber\\
    &+k^2[(\bm{S}_{h}\cdot\bm{R}_{ij})\{\bm{S}_{i}\times \bm{S}_{j}\cdot(\bm{R}_{ih}+ \bm{R}_{jh})\times\hat{z}\}+(\bm{S}_{h}\cdot\bm{R}_{ij}\times\hat{z})\{\bm{S}_{i}\times \bm{S}_{j}\cdot(\bm{R}_{ih}+ \bm{R}_{jh})\}]\frac{1}{8}(3\sin\phi-\sin3\phi),\\
    V_x&=S_h^z(\bm{S}_{i}\cdot \bm{S}_{j})(-\sin\phi+2\sin2\phi+\sin3\phi)\nonumber\\
    &+k^2S_h^z(\bm{S}_{i}\cdot \bm{S}_{j})\frac{1}{4}[R_{ij}^2(-5\sin\phi+4\sin2\phi-\sin3\phi)+\bm{R}_{ih}\cdot\bm{R}_{jh}(20\sin\phi-20\sin2\phi+8\sin3\phi-\sin4\phi)]\nonumber\\
    &+k^2(\bm{R}_{ih}\times \bm{R}_{jh}\cdot \hat{z})[\bm{S}_{h}\cdot\bm{S}_{i}\times \bm{S}_{j}(2\sin\phi-\sin2\phi)+S_h^z(\bm{S}_{i}\times \bm{S}_{j}\cdot \hat{z})\frac{1}{4}(4\sin\phi+2\sin2\phi-4\sin3\phi+\sin4\phi)]\nonumber\\
    &+k^2[(\bm{S}_{h}\cdot\bm{R}_{ij})\{\bm{S}_{i}\times \bm{S}_{j}\cdot(\bm{R}_{ih}+ \bm{R}_{jh})\times\hat{z}\}+(\bm{S}_{h}\cdot\bm{R}_{ij}\times\hat{z})\{\bm{S}_{i}\times \bm{S}_{j}\cdot(\bm{R}_{ih}+ \bm{R}_{jh})\}]\frac{1}{8}(-5\sin\phi+4\sin2\phi-\sin3\phi),\\
    V_y&=-ik\bm{S}_{h}\times(\bm{S}_{i}\times \bm{S}_{j})\cdot \bm{R}_{ij}(-3\cos\phi+4\cos2\phi-\cos3\phi),\\
    V_z&=-k(\bm{S}_{h}\cdot\bm{R}_{ij})(\bm{S}_{i}\times \bm{S}_{j}\cdot \hat{z})(5\sin\phi+4\sin2\phi+\sin3\phi)+kS_h^z(\bm{S}_{i}\times \bm{S}_{j}\cdot \bm{R}_{ij})(\sin\phi+2\sin2\phi+\sin3\phi).
\end{align} 
\end{widetext}
Among the asymmetric scattering terms, only those proportional to $\sin\phi$ contribute to Hall conductivity, as discussed in the Appendix.~\ref{sec:Boltzmann}.
Therefore, we focus on the terms proportional to $\sin\phi$, 
    \begin{align}
    \tilde{w}_{\bm{k}\mu\rightarrow \bm{k}'\nu}^-=\frac{27\pi |m| J_K^3 }{2^9(1-\alpha)L^4 }\tilde{V}_{\mu\nu}\sin\phi\delta\left(\epsilon_{\bm{k}\mu}-\epsilon_{\bm{k}'\nu}\right),
   \end{align}
where
\begin{align}
    \tilde{\bm{V}}=\tilde{V}_0\sigma_0+\tilde{V}_x\sigma_x+\tilde{V}_y\sigma_y+\tilde{V}_z\sigma_z,
\end{align}
and
\begin{widetext}
   \begin{align}
    \tilde{V}_0&=S_h^z(\bm{S}_{i}\cdot \bm{S}_{j})+k^2S_h^z(\bm{S}_{i}\cdot \bm{S}_{j})\frac{1}{4}[R_{ij}^2-4\bm{R}_{ih}\cdot\bm{R}_{jh}]+k^2(\bm{R}_{ih}\times \bm{R}_{jh}\cdot \hat{z})[2\bm{S}_{h}\cdot\bm{S}_{i}\times \bm{S}_{j}-S_h^z(\bm{S}_{i}\times \bm{S}_{j}\cdot \hat{z})]\nonumber\\
&+\frac{3}{8}k^2[(\bm{S}_{h}\cdot\bm{R}_{ij})\{\bm{S}_{i}\times \bm{S}_{j}\cdot(\bm{R}_{ih}+ \bm{R}_{jh})\times\hat{z}\}+(\bm{S}_{h}\cdot\bm{R}_{ij}\times\hat{z})\{\bm{S}_{i}\times \bm{S}_{j}\cdot(\bm{R}_{ih}+ \bm{R}_{jh})\}]\\
    \tilde{V}_x&=-S_h^z(\bm{S}_{i}\cdot \bm{S}_{j})+k^2S_h^z(\bm{S}_{i}\cdot \bm{S}_{j})\frac{5}{4}[-R_{ij}^2+4\bm{R}_{ih}\cdot\bm{R}_{jh}]+k^2(\bm{R}_{ih}\times \bm{R}_{jh}\cdot \hat{z})[2\bm{S}_{h}\cdot\bm{S}_{i}\times \bm{S}_{j}+S_h^z(\bm{S}_{i}\times \bm{S}_{j}\cdot \hat{z})]\nonumber\\
    &-\frac{5}{8}k^2[(\bm{S}_{h}\cdot\bm{R}_{ij})\{\bm{S}_{i}\times \bm{S}_{j}\cdot(\bm{R}_{ih}+ \bm{R}_{jh})\times\hat{z}\}+(\bm{S}_{h}\cdot\bm{R}_{ij}\times\hat{z})\{\bm{S}_{i}\times \bm{S}_{j}\cdot(\bm{R}_{ih}+ \bm{R}_{jh})\}],\\
    \tilde{V}_y&=0,\\
    \tilde{V}_z&=-5k(\bm{S}_{h}\cdot\bm{R}_{ij})(\bm{S}_{i}\times \bm{S}_{j}\cdot \hat{z})+kS_h^z(\bm{S}_{i}\times \bm{S}_{j}\cdot \bm{R}_{ij}).
\end{align} 
\end{widetext}
By using these equations and the semiclassical Boltzmann theory, the Hall conductivity $\sigma_{xy}$ for hole-doped case reads
\begin{align}
    \sigma_{xy}=-\frac{27\tau^2 e^2 |m| m J_K^3 \epsilon_F}{2^{11}\pi (1-\alpha)L^2}(\tilde{V}_0+\tilde{V}_x)|_{k=\sqrt{-2m\epsilon_F}},
\end{align}
where
\begin{align}
    \tilde{V}_0+\tilde{V}_x=k^2\left[f'_1(\{\bm{S}_{h}\})-4f_2(\{\bm{S}_{h}\})+\frac14f_3(\{\bm{S}_{h}\})\right].
\end{align}
The Hall conductivity for the electron-doped case was calculated in the same way as in the hole-doped case.
For the electron-doped case, the Hall conductivity reads
\begin{align}
    \sigma_{xy}=\frac{\tau^2 e^2 |m| m J_K^3 \epsilon_F}{2^{11}\pi (1+\alpha) L^2}(\tilde{V}_0+\tilde{V}_x)|_{k=\sqrt{2m\epsilon_F}},
\end{align}
where
\begin{align}
    \tilde{V}_0+\tilde{V}_x=-k^2\left[9f_1(\{\bm{S}_{h}\})+58f_2(\{\bm{S}_{h}\})+30f_3(\{\bm{S}_{h}\})\right].
\end{align}

\end{document}